\documentstyle[twoside,fleqn,espcrc2,epsf]{article}

\newcommand{\be}{\begin{equation}}
\newcommand{\ee}{\end{equation}}
\newcommand{\ba}{\begin{eqnarray}}
\newcommand{\ea}{\end{eqnarray}}

\title{Gauge-invariant strings in the 3d U(1)+Higgs theory}

\author{
K. Kajantie\address{Theory Division, CERN, CH-1211 Geneva 23, Switzerland},
M. Laine$^{\rm a}$,
T. Neuhaus\address{Helsinki Institute of Physics, P.O.Box 9, 
00014 University of Helsinki, Finland},
\underline{J. Peisa}\thanks{Poster presented by
J. Peisa.}\address{Department of Physics,
University of Wales Swansea, Singleton Park,
Swansea SA2 8PP, U.K.},
A. Rajantie$^{\rm b}$,
K. Rummukainen\address{NORDITA, Blegdamsvej 17,
DK-2100 Copenhagen \O, Denmark}
}

\begin{document}

\begin{abstract}
We describe how the strings, which are classical solutions of the
continuum three-dimensional U(1)+Higgs theory, can be studied on the
lattice. The effect of an external magnetic field is also discussed
and the first results on the string free energy are presented. It is shown
that the string free energy can be used as an order parameter when the
scalar self-coupling is large and the transition is continuous.
\end{abstract}

\maketitle

\section{Three-dimensional U(1)+Higgs theory}

The three-dimensional U(1)+Higgs theory
contains a complex scalar field $\phi(x) \equiv R(x) e^{i\varphi(x)}$
and an abelian gauge field $\alpha_i(x) \equiv eaA_i(x)$
and is
defined by the action 
\[
S = \int d^3x \left[ {1\over 4} F_{ij} F_{ij} + |D_i\phi|^2
          + m^2 \phi^*\phi + \lambda (\phi^*\phi)^2
          \right].
\]
This contains a dimensionful gauge coupling constant $e^2$, which
defines the scale of the system, and the two dimensionless parameters
\[
        x = {\lambda\over e^2} \mbox{ and }       y = {m^2\over e^4}.
\]
The continuum action can be discretised using either the compact or the
non-compact formalism for the gauge field -- in the following we shall
use exclusively the non-compact formalism, which has several advantages
over the compact one.

\section{Classical solutions} 

The three-dimensional U(1)+Higgs theory
contains classical, cylindrically symmetric solutions of the equations
of motion -- strings. These 
topological objects have been studied in great detail in connection with
superconductivity and cosmic strings. They can be characterized by
\[
\oint dx \nabla \varphi = 2\pi n_C,
\]
where $n_C$ is the {\em winding number}. This can be defined on the
lattice \cite{u1strings,plb} by 
computing the integer $n_C$ for a
closed loop $C$ by
\[
\sum_{l\in C}Y_l\equiv 2\pi n_C,
\]
where
\[
 Y_{({\bf x},{\bf x}+\hat i)}=[\alpha_i({\bf x})+\varphi({\bf x}+\hat i)-
\varphi({\bf x})]_\pi-\alpha_i({\bf x}).
\]

\section{External magnetic field}

The phase structure of the theory at zero external magnetic field is
strictly speaking only known in the compact theory \cite{u1higgs}. 
The study of the non-compact theory is in progress. However, it is known
that at small values of the scalar self-coupling~$x$, perturbation
theory works and the theory has a first order phase transition from the
symmetric to the broken phase. The transition weakens as $x$ is
increased. At some critical value $x_c \sim 0.5$ it is expected that
the transition becomes continuous. The continuous transition can be
observed 
by measuring the photon mass. We
suggest that in the non-compact theory the string free energy
could be used as an order parameter, as well.

The situation changes dramatically if an external magnetic field is
present. Depending on $x$, the system can have either two or four
phases. If $x$ is small, there is just one transition from the
symmetric to the broken phase (from normal to superconducting phase in
condensed matter terminology), while for large $x$ the system has also
a vortex phase. This vortex phase can be divided to two further
phases, a vortex lattice phase and a vortex liquid phase. The transition
between these two vortex phases has been observed to be first order in
real high $T_c$ superconductors \cite{realsc}, and it would be
interesting to see if this is predicted already by the Ginzburg-Landau
theory. At the moment, the computer simulations needed to answer this
question would be extremely costly.

A constant magnetic field in the $z$-direction corresponds to a
background gauge field, for instance $\alpha_i^{bg}=a e \delta_{i2} B
x$. However, the dynamics of the system may prefer to distribute the
flux of the magnetic field in an inhomogeneous way. In fact, the only
thing which remains constant and can be fixed is the total flux
through a given surface (e.g., the whole lattice).  On a lattice with
strictly periodic boundaries, there is no magnetic flux going through
the lattice. However, by choosing the boundary conditions to be, for
example,
\[
\alpha_2(N_x,0,n_z) = \alpha_2(0,0,n_z) + ea^2BN_xN_y,
\]
one forces a magnetic flux $ \Phi_0 = ea^2BN_xN_y$ to go
through the xy-plane. Note that this is only possible if one uses a
non-compact gauge field. Also, even though all values of the magnetic
field are possible in principle, values which lead to a non-periodicity
in the hopping term cause large inhomogeneities, so that in a finite
volume one can only use discrete values, namely $ea^2BN_xN_y=2\pi m$
for the magnetic field. Thus the allowed values of the flux are $
\Phi_0 = 2\pi m$.

Even though in principle the magnetic field cannot penetrate the
type I superconductor ($x$ small), the lattice construction above forces a
flux through the lattice. What happens is that a macroscopic volume
of the system remains in the symmetric phase, allowing the flux to
penetrate the lattice. In Type II superconductors it
is, however, expected that the magnetic field penetrates the system
forming real vortices. 

\section{\bf String density}

Even though the classical string solutions have a higher energy than
the true vacuum, it is possible that string-like objects are generated
by quantum or thermal fluctuations. This possibility has been studied 
in both the U(1)+Higgs theory \cite{plb} 
and the SU(2)+Higgs theory \cite{su2strings}.

In Ref. \cite{plb} the behavior of the string density -- number of
strings passing through a closed loop -- was studied. It was found
that at finite lattice spacing the $1\times 1$ loop shows clear
dependence of the parameters $x$ and $y$. However, at the continuum
limit the result is independent of the parameters (and can in fact be
obtained from a theory containing just a free massless scalar field)
-- the result is pure UV noise.

It was also shown that the $\beta_G \times \beta_G$-loop, which has a
constant physical size as the lattice spacing is decreased, can
contain divergent parts or can get a contribution from the
UV-noise. The easiest way to remove unphysical contributions is to
study the differences of string densities at various $y$, 
which should be free of the UV-noise.

\section{Vortex Free Energy}

We present a study of the vortex free energy $T$ per unit length 
at $\beta_{G}=4$ and at $x=2$. This observable is an order parameter, 
as its value is finite in the Higgs phase and zero in the Coulomb phase.

A quantized magnetic flux of magnitude $\Phi=2\pi$ in a finite box
is forced into the system in one of the arbitrary chosen lattice
directions, as described above. The gauge part of the action on the 
lattice then has the form 
\[ 
S_{\rm gauge}={\beta_G \over 2} \sum_{P} (\alpha_P-2\pi
\bar{\alpha}_P)^2,
\] 
otherwise the theory remains unchanged. Here the $\bar{\alpha}_P$ have
nonvanishing values $\bar{\alpha}_P=+1,-1$ only, if their dual coincides with 
a closed loop, which on the torus is closed by the boundary. 

In the Higgs phase (at small values of $y$) the vortex free energy $T$ per
unit length is finite,
\[ 
T= {1\over L} \ln{Z(\Phi=0)/Z(\Phi=2\pi)}.
\]
In this phase the magnetic flux is confined into a string and a
dislocation of linear dimension $L$ is formed in the $L^3$ boxes
considered.

In the Coulomb phase (large values of $y$) the vortex free energy $T$ 
vanishes.  Here the theory exhibits long range order as characterized by a 
massless excitation. Large finite size effects within $T$ signal 
the existence of photons.

\begin{figure}
\vspace{-2.0cm}
\leavevmode
\hspace{-0.5cm}
\epsfysize=10cm
\epsffile{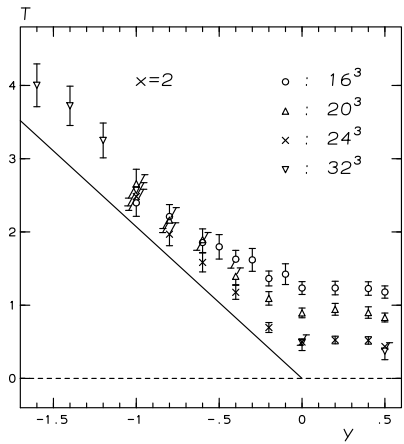}
\vspace{-2.5cm}
\caption{The vortex free energy $T$ per unit length plotted as a function 
of $y$ at  $x=2$. The solid line corresponds to the mean field result.}
\vspace{-0.5cm}
\end{figure}

\begin{figure}
\vspace{-2.0cm}
\leavevmode
\hspace{-0.5cm}
\epsfysize=10cm
\epsffile{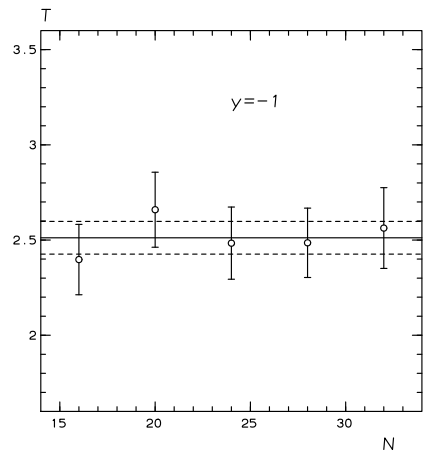}
\vspace{-2.5cm}
\caption{The vortex free energy for $x=2$, $y=-1$ as a function of the 
linear extent $L=Na$. There are no visible finite size effects.}
\end{figure}

In Figures~1 and 2 we present Monte Carlo measurements of the
quantity $T$. It can been seen that $T$ is finite
in the small y-region of the theory. Finite size effects in the quantity 
$T$ are under control at $y=-1.0$, as can be seen from Figure~2.

There are large finite size effects in the quantity $T$ for
values of $y \ge 0$ (see Fig.~1). These finite volume values of
$T$ extrapolate to the value $0$ at positive values of $y$. The vanishing of 
$T$ there and the observation of large finite size
effects is consistent with the presence of a massless mode.

Our data in the Higgs region are close to the mean field result. 
Mean field scaling is expected to be valid away from the critical point.
It is an open question, whether the critical singularity is dominated
by mean field exponents or not. 
A more conclusive study is currently being prepared~\cite{tension}.

JP acknowledges the financial contribution of the European Commission
under the TMR-Programme ERBFMRX-CT97-0122.

\end{document}